\documentclass[twocolumn,preprintnumbers,amsmath,amssymb,superscriptaddress,nofootinbib,longbibliography]{revtex4-1}

\usepackage[english]{babel}
\usepackage[utf8]{inputenc}
\usepackage{amssymb}
\usepackage{amsmath}
\usepackage{color}
\usepackage{hyperref}
\usepackage{bbm}
\usepackage{epsfig}
\usepackage{multirow}
\usepackage[dvipsnames]{xcolor}
\usepackage{ulem}

\newcommand{\e}{\varepsilon}
\newcommand{\s}{\sigma}

\newcommand{\up}{\uparrow}
\newcommand{\down}{\downarrow}
\newcommand{\w}{\omega}

\newcommand{\de}{{\rm d}}
\newcommand{\dd}{\partial}
\newcommand{\Pol}{\mathcal{P}}
\newcommand{\T}{\mathcal{T}}

\newcommand{\mean}[1]{\left\langle {#1} \right\rangle }
\newcommand{\GF}[1]{  \langle\!\langle #1 \rangle\!\rangle  }
\newcommand{\vk}{\mathbf{k}}

\newcommand{\GS}[1]{ \Gamma_{{\rm S}#1} }
\renewcommand{\Im}{\mathrm{Im}}

\newcommand{\dk}{d^\dagger}
\newcommand{\ck}{c^\dagger}

\newcommand{\beq}{ \begin{equation} } 
\newcommand{\eeq}{ \end{equation} }
\newcommand{\beqa}{\begin{eqnarray}}
\newcommand{\eeqa}{\end{eqnarray}}
\newcommand{\nn}{\nonumber}
\newcommand{\es}{& = &}

\newcommand{\fig}[1]{Fig.~\ref{fig:#1}}
\renewcommand{\sec}[1]{Sec.~\ref{sec:#1}}
\newcommand{\eq}[1]{Eq.~(\ref{#1})}

\hypersetup{
    bookmarks=false,        
    colorlinks=true,        
    linkcolor=red,        
    citecolor=blue,        
    filecolor=blue,      
    urlcolor=blue
}

\begin{document}

\title{Interplay of the Kondo effect with the induced pairing in electronic and caloric
properties of T-shaped double quantum dots}

\author{Krzysztof P. W{\'o}jcik}
\email{kpwojcik@ifmpan.poznan.pl}
\affiliation{Institute of Molecular Physics, Polish Academy of Sciences, 
			 Smoluchowskiego 17, 60-179 Pozna{\'n}, Poland}
\affiliation{Faculty of Physics, Adam Mickiewicz University, 
			 Umultowska 85, 61-614 Pozna{\'n}, Poland}

\author{Ireneusz Weymann}
\affiliation{Faculty of Physics, Adam Mickiewicz University, 
			 Umultowska 85, 61-614 Pozna{\'n}, Poland}

\date{\today}


\begin{abstract}
We examine the influence of the superconducting proximity effect on the transport properties of a T-shaped 
double quantum dot strongly coupled to two normal, nonmagnetic or ferromagnetic leads. We show that the
two-stage Kondo screening may be suppressed or enhanced by the presence of pairing correlations,
depending on the specific geometric arrangement of the device.
We explain our results by invoking effective decrease
of Coulomb interactions by proximity effect and find qualitatively correct description 
in many cases, although spin-filtering effect stemming from spin-dependent Fano-Kondo interference 
occurs to be surprisingly fragile to the presence of induced superconducting pairing correlations.
The results are obtained within the numerical renormalization group framework in the limit of large 
superconducting gap, which allows for a reliable examination of
the low-temperature sub-gap properties of the considered system.
Nevertheless, finite temperature effects are also taken into account. 
\end{abstract}

\maketitle

\section{Introduction}
\label{sec:intro}

The two-stage Kondo effect in double quantum dots (DQDs) or double magnetic impurities 
has been studied for over a decade, both theoretically 
\cite{Pustilnik, Vojta, vanderWiel_PRL02, Cornaglia, ZitkoBonca, ZitkoPRB2010, 
Ferreira, TanakaPRB2012, KWIW-2stageKondo, KWIW-termo}
and experimentally \cite{Craig2004_Science04,Granger_PRB05}, in various contexts.
In particular, its relation to the Fano-like interference \cite{Fano, T-shapedFano, Sasaki}
was precisely established \cite{ZitkoPRB2010} and the spin-dependent variant of this effect
for DQDs in an external magnetic field \cite{daSilva}
or coupled to ferromagnetic leads \cite{KWIW-P} was proposed as a method for obtaining 
electrically tunable spin-polarized currents.
Moreover, the Andreev transport properties of 
T-shaped DQDs coupled to superconducting (SC) and normal leads
have also been considered \cite{TanakaPRB2008,JBTD,Baranski_PRB12,Calle_PLA13,Baranski_CP15,Calle_JPCM17}.
In such hybrid systems, for low temperatures
and voltages smaller than the superconducting energy gap,
transport occurs due to Andreev reflection
processes \cite{andreev,DeFranceschi_NatNano10,Martin-Rodero_AP11}.
However, while most studies dealt with transport between normal and superconducting
electrodes, the normal electronic and caloric transport through T-shaped DQDs coupled 
to two normal (ferromagnetic) leads and proximized by the third,
superconducting electrode, has been hardly examined so far. 
Therefore, in this article we perform a detailed 
and accurate analysis of such a case.

To begin with, it is instructive to notice that similar studies of
a single quantum dot case unveiled an intriguing interplay between the Kondo 
physics \cite{Kondo,Hewson} and the pairing induced by the superconducting contact
\cite{RozhkovArovas,Buitelaar,Franke_Sci11,Pillet_PRB13}.
A hallmark of this interplay is a quantum phase transition
between the Kondo-screened singlet and the BCS-like
singlet states, as the ratio of the Kondo temperature to
the superconducting energy gap is varied \cite{Franke_Sci11,Pillet_PRB13}.
At the Kondo side of the transition, the Kondo temperature was found to be enhanced with 
increasing the coupling strength to the superconducting lead \cite{KWIW,DomanskiIW}.
At the other side, Yu-Shiba-Rusinov-like bound states \cite{yu,shiba,rusinov} are formed, 
which have already been explored experimentally with the Andreev bias spectroscopy \cite{Lee_NatNano14}. 
The quantum phase transition is present only in the absence of the normal leads
and gets smeared to a crossover otherwise. Nevertheless, even in the latter case, 
around the critical value of the quantum-dot---SC coupling, the BCS-like expectation 
value $\mean{d_\up d_\down}$ for spin-$\s$ quantum dot annihilation operators $d_{\s}$ 
becomes nonzero \cite{DomanskiIW}.

In this paper we show that the interplay of superconducting proximity effect
and correlations giving rise to the Kondo effect is even more interesting,
if a single quantum dot is substituted by a T-shaped DQD; see \fig{system}.
In this geometry one quantum dot (QD1 in \fig{system})
is embedded between two normal (ferromagnetic) leads and coupled to 
the second quantum dot (QD2). We consider two possible scenarios,
in which the superconductor is coupled to either the first or the second quantum dot. 
Then, depending on which quantum dot is proximized, and what is the strength of the
coupling to SC lead, different interesting effects take place, as described in the following.
They include, among others, an enhancement and a destruction of any of the two screening stages
of the Kondo phenomenon.

Since one of the most experimentally accessible physical quantities
of such a system is its conductance,
we base our discussion on the dependence of conductance
on model parameters, gate voltages and temperature.
This allows us to thoroughly examine the
influence of induced superconducting pairing
on the two stages of the Kondo effect.
Moreover, further information is gained from the analysis of the Seebeck coefficient,
the so-called thermopower, whose sign and magnitude change between different Kondo states
\cite{KrawiecWysokinski, CostiZlatic, Zitko_QD+B, Zitko_overscreened, ZitkoBulka, Donsa, KWIW-termo},
while temperature dependence allows for recognizing metallic and hopping-like transport regimes
\cite{Mott,ZitkoBulka}.

We note that the subgap transport 
through hybrid double quantum dot systems is currently undergoing
an extensive exploration. This has been stimulated by
impressive experiments demonstrating controllable splitting
of Cooper pairs in DQDs with both dots attached to a superconductor \cite{CPS}.
This has also provided a great motivation to many researchers to
analyze hybrid DQDs in terms of their Cooper pair splitting efficiency.
This is an undoubtedly interesting direction, however, here we focus on 
completely different geometry, with only one quantum dot directly coupled to the SC lead.
While being less useful as a Cooper pair splitter,
this system exhibits very interesting strongly-correlated physics.
Simultaneously, recent rapid experimental advances in the field 
\cite{CPS2,CPS3,CPS4,CPS5,CPS6,CPS7,CPS8,Grove-Rasmussen2017Nov}
give hope for a possibility of fabricating the device considered here.
From this point of view, our results are expected to
stimulate further research in hybrid T-shaped DQD
as well as to be of assistance in understanding future experimental observations.

It is interesting to note that the interplay
between the Kondo correlations and superconductivity
has been also considered in the case of 
the Anderson model with attractive on-site Coulomb interactions \cite{Anderson_negU}.
In such a case the charge Kondo effect may occur, manifesting itself in the electronic \cite{Coleman_negU},
caloritronic \cite{Andergassen_negU} and spin-caloritronic \cite{IW_negU} properties.
Moreover, intensive theoretical and experimental investigations have clearly shown that in Tl-doped PbTe
the negative-$U$ centers induce superconductivity in the otherwise normal host, while the charge
Kondo effect takes place in the system \cite{Dzero_negU,Fischer1,Fischer2,Fischer3,CostiZlatic_negU}.
The charge Kondo effect is, however, not present in our system. Instead of attractive-$U$ 
center influence on the normal host, we examine the influence of BCS superconductor on a
double quantum dot structure.
Furthermore, recent experiments have also demonstrated
the possibility of fabricating quantum dots with attractive 
Coulomb interactions, which persist both below and above the critical temperature for the 
superconducting transition in the leads \cite{Cheng_Nature,Cheng_PRX}. This gives rise
to an interesting interplay between the electrostatic attraction and pairing, which leads to suppression of 
the super-current through the device in the crossover region between the weak-coupling and 
strong-coupling unitary transmission regimes \cite{Fang1}.
Moreover, unlike the spin Kondo effect, its charge counterpart
may become enhanced under nonequilibrium spin bias \cite{Fang2}.
Although in this paper we focus on the repulsive $U$ case,
our work shall contribute to the general understanding 
of the interplay between Kondo correlations with the superconducting proximity effect.

The paper has the following structure. In \sec{model} a detailed description of the model is provided.
Section \ref{sec:U} briefly summarizes the role of the magnitude of Coulomb interactions for further reference.
The results for the case of QD1 (QD2) coupled to the SC lead are then presented in \sec{QD1-SC}
(\sec{QD2-SC}), respectively, and the paper is summarized in \sec{conclusions}. 

\section{Model}
\label{sec:model}

\begin{figure}[bt]
\centering
\includegraphics[width=0.85\columnwidth]{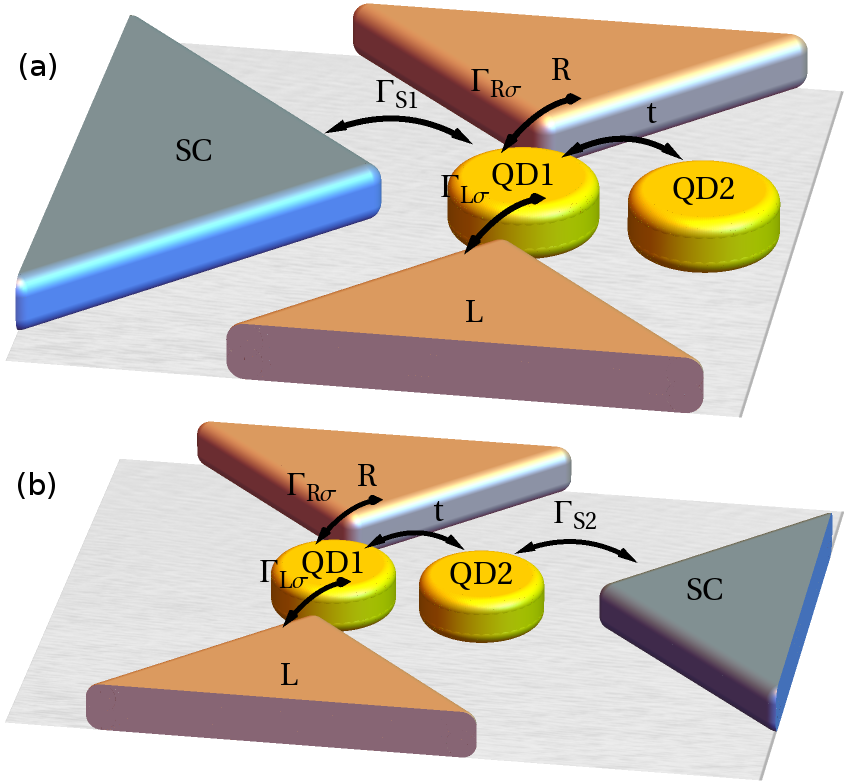}
\caption{Schemes of possible realizations of the considered system.
The first quantum dot (QD1) is coupled to the left and right
normal (ferromagnetic) leads with coupling strength $\Gamma_{r\s}$,
where $r=L/R$ for the left/right lead. The two dots are coupled 
via hopping matrix elements $t$.
The superconducting electrode can be attached
either to (a) the first ($i=1$)
or to (b) the second ($i=2$) quantum dot, with
the corresponding coupling strength $\Gamma_{Si}$.}
\label{fig:system}
\end{figure}

In the present paper we consider the T-shaped double quantum dot (DQD) coupled to two 
metallic (in general ferromagnetic) leads, and proximized by one superconducting electrode.
We analyze two possible realizations of such system,
in which the SC lead is attached either to the first [\fig{system}(a)]
or to the second [\fig{system}(b)] quantum dot.
In both cases, the Hamiltonian of the system can be written in the general form 
$H = H_{\rm DQD} + H_{\rm L} + H_{\rm R} + H_{\rm T} + H_{\rm S} + H_{\rm TS}$,
where the subsequent parts describe the isolated DQD, left and right leads, 
tunneling between DQD and these leads, superconductor, 
and finally the tunneling between SC and DQD, respectively.

We assume that the normal leads contain quasi-free electrons, 
$H_{r} = \sum_{\vk\s} \e_{r\vk\s} \ck_{r\vk\s} c_{r\vk\s}$, with $r\in\{{\rm R},\rm{L}\}$
and $c_{r\vk\s}$ denoting the annihilation operator corresponding to electron in lead $r$ possessing 
pseudo-momentum $\vk$ and spin $\s$.
$H_{\rm T}$ has a form of spin-preserving local hopping between QD1 and the electrodes, 
$H_{\rm T} = \sum_{r\vk\s} v_{r\vk} \dk_{1\s} c_{r\vk\s}$, where $d_{i\s}$ annihilates spin-$\s$ electron
at QD$i$. Assuming the wide-band situation,
for the hybridization function between QD1 and lead $r$ we take a constant within the energy cut-off 
$\pm D$ around the Fermi level, $\Gamma_{r\s} = \pi \rho_{r\s} |v_{kr}|^2$, where $\rho_{r\s}$ is the 
(spin-resolved) normalized density of states in the lead $r$ at the Fermi level. 
With these approximations ferromagnetism of normal leads can be taken into account through the
spin dependence of $\Gamma_{r\s} = \Gamma_r(1+p_r\s)$, where $p_r$ is the spin polarization in the 
lead $r$, provided their magnetization is parallel. We also assume symmetric couplings, 
$\Gamma_r = \Gamma/2$, and $p_{\rm L} = p_{\rm R} = p$.

In the present paper we focus on the low-temperature physics.
Therefore, having written $H_{\rm S}$ in the BCS form, 
$H_{\rm S} = \sum_{\vk\s} \e_{{\rm S}\vk} \ck_{{\rm S}\vk\s}c_{{\rm S}\vk\s} 
+ \sum_{\vk} (\Delta_{\vk} \ck_{{\rm S}\vk\up}\ck_{{\rm S}-\vk\down} + h.c.)$,
we assume isotropic pairing amplitude, $\Delta_{\vk} = \Delta > 0$,
and integrate out the single-electron states 
of the superconductor lying outside the energy gap $2|\Delta|$, to finally take the limit
of $|\Delta|\to\infty$ \cite{RozhkovArovas,Buitelaar}.
In this way we obtain an effective Hamiltonian
$H_{\rm eff} = H_{\rm SDQD} + H_{\rm L} + H_{\rm R} + H_{\rm T}$,
with SC-proximized DQD part
\beqa
H_{\rm SDQD} 	\es 
				\sum_{i\s} \e_{i} n_{i\s} 
				+\sum_{i} U_i n_{i\up} n_{i\down} 
				+U' (n_1-1)(n_2-1) 
				\nn\\&&
				+t \sum_\s (\dk_{1\s}d_{2\s} + h.c.) 
				-\GS{i} (\dk_{i\up} \dk_{i\down} + h.c.),
	\label{H_DQD} 
\eeqa
where $\e_i$ is the energy level of QD$i$, $U_i$ denotes the respective Coulomb interaction strength, 
$U'$ measures inter-dot Coulomb interactions, $t$ is the inter-dot hopping matrix element and
$\GS{i}$ describes the coupling to the superconductor of QD$i$ ($i=1$ or $2$, depending on geometry). 
The operator $n_i = n_{i\up} + n_{i\down}$, while $n_{i\s} = \dk_{i\s}d_{i\s}$.
Henceforth, we use the detuning $\delta_i= \e_i+U_i/2$ from the particle-hole symmetry point of 
each dot to specify the energy levels of QDs. The coupling $\GS{i}$ is related to the hopping 
matrix element $v_{{\rm S}i}$ between QD$i$ and SC, and the normalized density of states of SC 
in the normal state, $\rho_{\rm S}$, through $\GS{i} = \pi \rho_{\rm S} |v_{{\rm S}i}|^2$, 
and it is assumed to be energy-independent, similarly to the normal leads case.
The negative sign in front of $\GS{i}$ corresponds to the choice of real and positive
$\Delta$ in the bulk superconductor Hamiltonian. 
The second quantum dot, QD2, is by assumption coupled to the normal leads
only indirectly, through QD1; compare \fig{system}.
Through an even-odd change of basis of the leads states \cite{even-odd},
the model at equilibrium can be exactly mapped onto an effective single-band system,
possessing an effective coupling $\Gamma$ and a spin polarization $p$. 

Then, the model is solved with the aid of the numerical renormalization group (NRG) technique
\cite{WilsonNRG,fnrg}. We use the complete basis set \cite{AndersSchiller1,AndersSchiller2}
to construct the full density matrix of the system \cite{Weichselbaum}. Once the energy
spectrum of the discretized Hamiltonian is known, the spin-resolved transmission coefficient 
$\mathcal{T}_\s(\w) = - \Gamma_\s \Im \GF{d_{1\s}; \dk_{1\s}}^{\rm ret.}(\w)$ is calculated 
from the imaginary part of the Fourier transform of the retarded QD1 Green's function.
The transport coefficients, such as the linear-response conductance 
in spin-channel $\s$, $G_\s$, and the thermopower, $S$, can be calculated from $\mathcal{T}_\s(\w)$ 
using the standard linear-response expressions
\beqa
G_\s \es \frac{e^2}{h} L_{0\s} , 
	\label{G} 
	\\
S \es - \frac{1}{eT} \frac{L_{1\up}+L_{1\down}}{L_{0\up}+L_{0\down}},
	\label{S}
\eeqa
with $L_{n\s} = \sum_\s \int \w^n [-\dd f_T(\w)/\dd\w] \mathcal{T}_\s(\w) \de\w$, $f_T(\w)$ denoting the 
Fermi-Dirac distribution function, $e$ (minus) the electron charge, and $h$ the Planck constant. 
The spin-dependent conductance allows for determining
the linear-response current spin polarization through
$\Pol = (G_\up-G_\down)/G$, with the total conductance $G=G_\up + G_\down$.
In NRG calculations at least $2048$ states per iteration were kept,
the discretization parameter $\Lambda=2$ was used,
while the quantities of interest were calculated directly
from discrete data \cite{weymann_PRB13}.

While neglecting the presence of the states of the superconductor lying outside the gap is one of the
strongest limitations of the presented model, one needs to keep in mind that at low temperatures
these states contribute quite weakly to the physics of the real systems. Moreover, the device is 
coupled to another continuum, namely to normal leads. Therefore, one can expect that the effects
of the presence of gapped continuous part of the spectrum of SC lead are only quantitative and rather 
weak at low temperatures. 
Nevertheless, detailed study of a single quantum dot coupled to superconductor \cite{OguriHewson}
show the sign change of the order parameter at the singlet-doublet transition point, which is 
necessarily not captured in our model for the quantum dot directly coupled to the SC electrode.

\section{The role of Coulomb interactions}
\label{sec:U}

One of the most intuitive consequence of the presence of a pairing potential induced by 
SC proximity is an effective reduction of the corresponding Coulomb repulsion. To be able to analyze 
the range of validity of this picture, first we summarize the effects related to the on-dot
and inter-dot capacitative correlations, $U_i$ and $U'$, for further reference. Therefore, 
in this section we consider the system in the absence of SC lead.

\subsection{Influence on Kondo screening}
\label{sec:Kondo_U}

\begin{figure}
\includegraphics[width=0.9\linewidth]{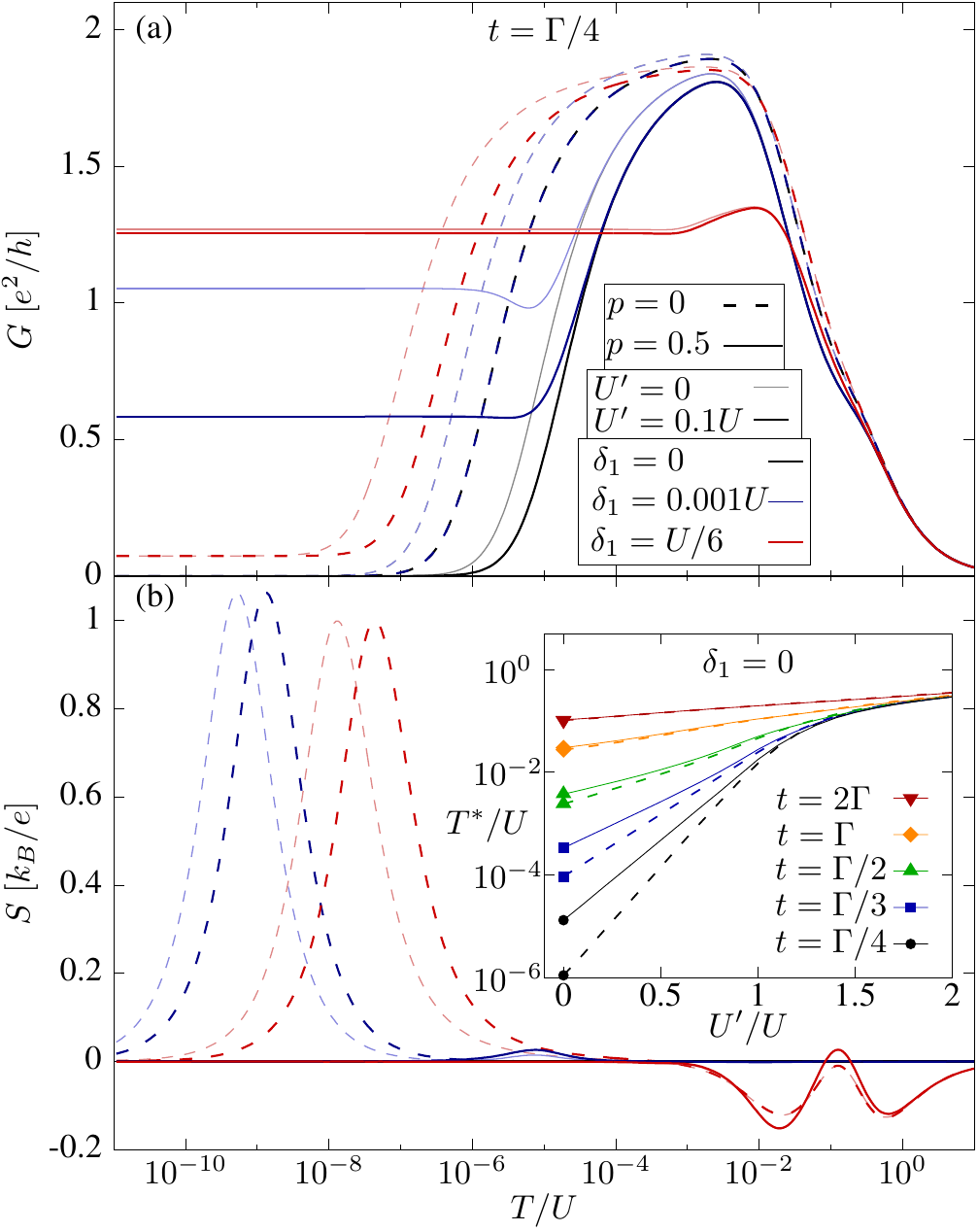}
\caption{(a) The conductance $G$ and (b) the Seebeck coefficient $S$
		 as functions of temperature $T$ for different detunings $\delta_1$, $U_1=U_2=U=D/10$,
		 $\Gamma=U/5$ and $t=\Gamma/4$. Here $D$ is the band halfwidth
		 used as energy unit. Solid lines correspond to finite spin polarization 
		 of the leads, $p=0.5$, while dashed
		 lines were used for $p=0$. Thick (bright thin) lines indicate the presence 
		 (absence) of inter-dot Coulomb interaction $U'=U/10$. 
		 The inset shows the dependence of the second-stage Kondo temperature
		 $T^*$ on $U'$ for the particle-hole symmetric case and different $t$.
		 }
\label{fig:Kondo_vs_U}
\end{figure}

The essence of the Kondo effect is the screening of the local moment by the conduction
band electrons \cite{Kondo}. Since Coulomb interactions are inevitable for the formation
of such a moment, they are clearly necessary for the Kondo physics to occur. However,
it should be also noted that for $U \gtrsim 4\Gamma/\pi$, which is the most common situation, 
the Kondo temperature $T_K$ is a decreasing function of $U$ due to its exponential 
dependence on $\Gamma/U$ \cite{Haldane}. 

In T-shaped DQDs the Kondo effect develops in two stages \cite{Cornaglia}. 
When the temperature is lowered, first,
the magnetic moment of QD1 is screened by the conduction electrons of the leads
at the Kondo temperature $T_K$. Then, for $T \ll T_K$, the resulting Fermi liquid serves
as a band of the half-width $\sim T_K$ for the second quantum dot (QD2),
the magnetic moment of which is screened at the second stage of the Kondo effect,
with the corresponding Kondo temperature \cite{Cornaglia}
\beq
T^* \sim T_K \exp(-T_K/J),
\label{Tstar}
\eeq 
where $J$ is an effective antiferromagnetic exchange
interaction between the two dots, $J\sim t^2/U$.
Note that estimations of $T_K$ or $T^*$, such as \eq{Tstar}, possess rather an
order-of-magnitude precision and for qualitative comparison of Kondo temperatures 
in different systems more precise definition is necessary. Here, we follow the
convention of defining $T_K$ as a temperature at which the conductance 
increases to half of its maximal value as the temperature is lowered,
such that $G(T_K) = G_{\rm max}/2$,
with $G_{\rm max}$ being the global maximum of $G(T)$.
Moreover, in this paper by $T_K$ we mean in fact the Kondo temperature in the case of $t=0$. 
Furthermore, in a similar fashion we can define $T^*$
as the temperature below $T_K$ at which $G(T)$ drops to $G_{\rm max}/2$
again (this happens only for $t\neq 0$).

The picture of the two-stage screening presented above
does not include the influence of capacitive coupling between the
two dots, $U'$, which will be discussed now.
Figure \ref{fig:Kondo_vs_U} demonstrates how finite values of $U'$ influence
the Kondo physics in the considered nanostructure,
depending on detuning of QD1 $\delta_1$ from the particle-hole symmetry (PHS) point, $\delta_1=0$, 
in the case of nonmagnetic ($p=0$) and ferromagnetic ($p=0.5$) metallic leads.
In \fig{Kondo_vs_U}(a) one can see that the second-stage Kondo temperature
$T^*$ is indeed increased by finite $U'$.
In fact, the effective exchange coupling $J$ increases by a factor $(1-U'/U)^{-1}$ for finite
capacitive coupling between the dots \cite{Ferreira}.
However, qualitative features remain the same. At PHS, with lowering the temperature,
the conductance first increases at $T_K$
and almost reaches $2e^2/h$.
Then, it decreases to $0$ for temperatures below $T^*$. This behavior 
is observed for both ferromagnetic and nonmagnetic leads, although
only at the PHS point. There, the role of the leads' spin polarization $p$
is reduced to a change in $T_K$ \cite{MartinekTK} and, thus, the following 
change in $T^*$, cf.~Eq.~(\ref{Tstar}).

A small detuning from the PHS point results in only quantitative changes for $p=0$, yet 
it completely changes the situation for finite $p$. As clearly visible in \fig{Kondo_vs_U}(a),
$G(T)$ does not drop to $0$ at low temperatures for finite $\delta_1$. However, 
the residual conductance is quite small even for relatively large detunings in
the case of $p=0$, while for finite $p$, the conductance remains large at low $T$. 
This is caused by the exchange field induced by the ferromagnetic leads 
\cite{MartinekEx,KWIW-2stageKondo}.
This exchange field strongly depends on the position of the quantum dot levels
and vanishes precisely at the PHS point
\cite{MartinekEx,KWIW-2stageKondo}.
Once the exchange field becomes larger than $T^*$ (which is in fact very small), the second stage 
of the Kondo effect is blocked and the conventional ({\it i.e.} single-stage)
Kondo effect is restored. On the other hand, for large detunings
[compare the curve for $\delta_1=U/6$ in \fig{Kondo_vs_U}(a)],
the exchange field is comparable to $T_K$ and also the conventional Kondo effect 
becomes blocked.

In the inset of \fig{Kondo_vs_U} the dependence of $T^*$ on $U'$ is presented
for a few values of $t$ and $p=0$ (dashed lines) as well as $p=0.5$ (solid lines). 
It was extracted from $G(T)$ dependences calculated for different $U'$.
As reported earlier by Ferreira and co-workers for the case of nonmagnetic leads \cite{Ferreira}, 
the capacitative coupling between the dots tends to increase $J$ and leads to 
exponential increase of $T^*$ in the physically relevant regime of $U'<U$.
This remains true also for ferromagnetic leads. Actually, the presence of 
Coulomb correlations between the dots
reduces the difference between the cases of finite $p$ and $p=0$,
which is an interesting result at PHS point, where the only influence of $p$ is the $T_K(p)$ dependence.

Additional information about the relevant regimes can be extracted from the temperature
dependence of the thermopower $S$ \cite{CostiZlatic,KWIW-termo}.
However, to achieve finite values of the Seebeck coefficient,
one needs to tune the system from the PHS point, where $S=0$.
Let us now inspect this in more detail for the line corresponding to $p=0$,
$\delta_1=U/6$ and $U'=U/10$ shown in \fig{Kondo_vs_U}(b).
At high temperatures the system is in the hopping transport regime \cite{Mott,ZitkoBulka},
characterized by $S\sim T^{-1}$. Negative sign of $S$
is caused by the fact that positive frequencies host
more spectral weight. Then, with decreasing the temperature,
$S$ exhibits first a local minimum and then, while cooling the system further,
its sign changes twice, before another minimum occurs. The narrow 
region of positive thermopower corresponds to the Coulomb blockade regime,
which is hardly present due to relatively strong coupling $\Gamma=U/5$
used in \fig{Kondo_vs_U}. The second minimum in $S$ is a consequence
of asymmetric Kondo peak near the Fermi level. Despite the fact that $T_K$ depends on $p$ 
\cite{MartinekTK}, the position of the minimum related to the Kondo effect is
practically independent of $p$. Moreover, it also hardly depends on $U'$; cf. \fig{Kondo_vs_U}(b). 
This is not the case for the position of the maximum in thermopower,
which is present at even lower temperatures and is related to
the second stage of screening.
One can also see that the maximum is completely absent for $p=0.5$, which is due to the fact
that for assumed parameters the exchange field is larger than $T^*$
and the second stage of screening is suppressed;
compare with \fig{Kondo_vs_U}(a).
Furthermore, as far as the effect of $U'$ is concerned,
the shift of the maximum in $S$
due to capacitive coupling can be visible and it 
results from the corresponding change in $T^*$,
which can be seen in the temperature dependence of the conductance.

Finally, it is worth to note that the maximum of $S$ at $T\sim T^*$ is much more pronounced
as compared to the minimum at $T\sim T_K$. This is caused by the fact, that good thermoelectric materials 
are characterized by sharp and asymmetric features in the spectral density near $\w=0$ 
\cite{Hicks_Dresselhaus,MahanSofo}. For the parameters considered in \fig{Kondo_vs_U},
the Kondo temperature $T_K$ is quite large and
the Kondo peak in the spectral density is relatively broad. On the contrary,
$T^*$ is indeed cryogenic, and the dip in $\T(\w)$ corresponding
to the second stage of screening is very sharp.

\subsection{Influence on Fano interference and its spin dependence}
\label{sec:Fano_U}

\begin{figure}
\includegraphics[width=0.9\linewidth]{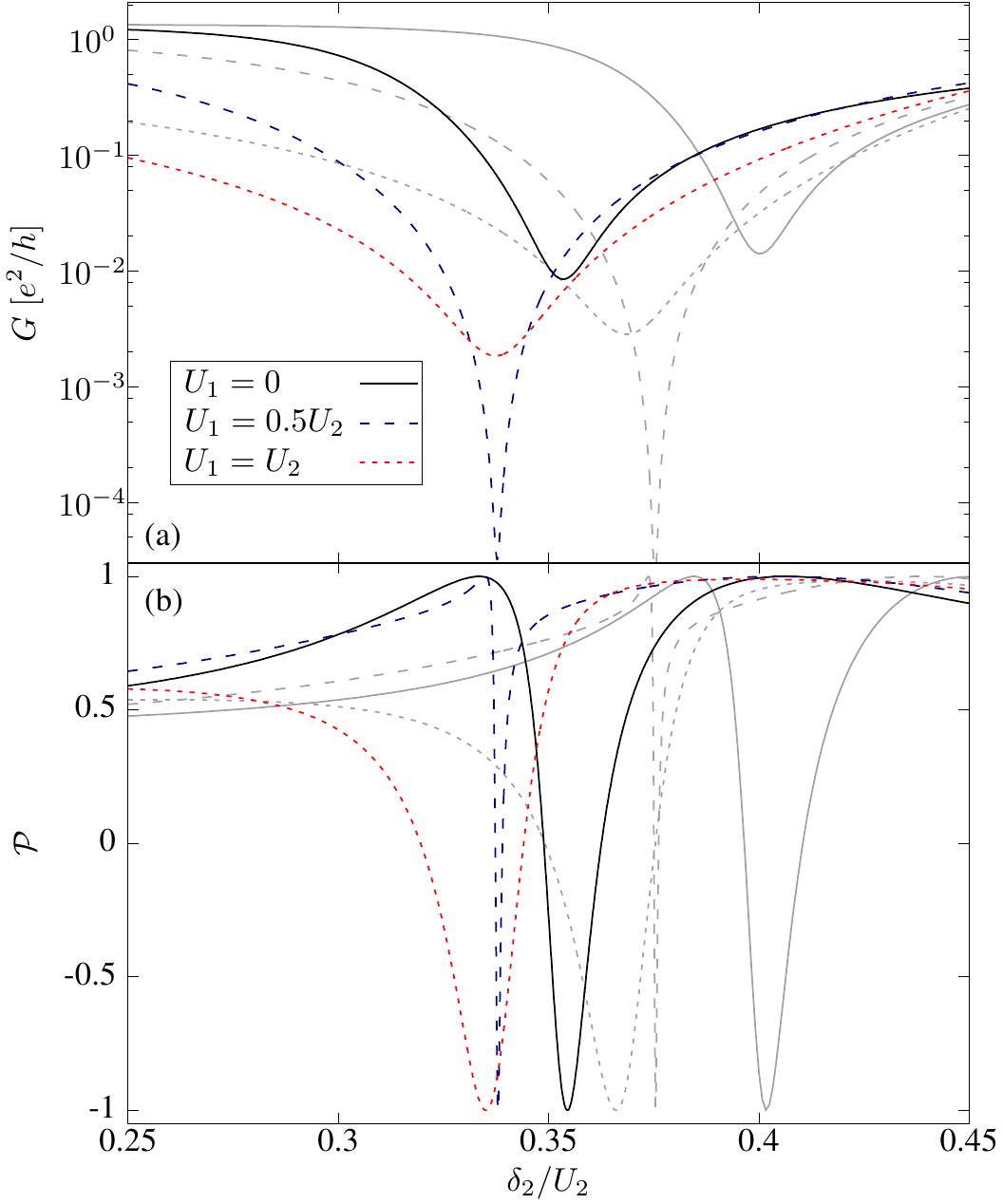}
\caption{(a) The low-temperature conductance $G$ and (b) its spin polarization $\Pol$
         plotted as a function of QD2 detuning $\delta_2$ for ferromagnetic leads ($p=0.5$), 
         $\delta_1 =-0.1U_2$, $U'=0.1U_2$, $U_2=D/10$ and different $U_1$, as indicated.
		 For comparison, the same curves calculated for the case of $U'=0$ 
		 are shown using a light grey color.
		 }
\label{fig:Fano_vs_U1}
\end{figure}

The Fano effect is a consequence of the quantum interference between
a resonant level and the continuum of states \cite{Fano}.
It is therefore also present in DQD systems (even noninteracting) and manifests itself
through an antiresonance in the conductance as a function of DQD energy levels
\cite{T-shapedFano}.
Finite Coulomb correlations can modify the conditions for Fano interference
and result in another interesting phenomena. 
Primarily, the Fano physics is obtained only at the zero-temperature limit, which may be experimentally
irrelevant due to cryogenic scale of $T^*$ occurring in the system. At finite $T$, deviations from Fano 
anti-resonance curve can be expected and has already been measured \cite{ZitkoPRB2010,Sasaki}.
In fact, the antiresonance itself may be seen as the consequence of the second stage of the Kondo effect,
which leads to the suppression of the conductance at $T \ll T^*$ \cite{ZitkoPRB2010}. Moreover,
when $U_2>0$, a spin-splitting of the conductance antiresonance
occurs in T-shaped DQD coupled to ferromagnetic leads
without applying an external magnetic field  \cite{KWIW-P}.

The Fano-like antiresonance is visible in \fig{Fano_vs_U1}(a), where the conductance is plotted 
against detuning of QD2 energy level for a few values of the Coulomb interaction strengths of QD1, 
$U_1$. Clearly, for all considered values of $U_1$, the minimum in $G(\delta_2)$ is present (note the logarithmic
scale on vertical axis). The total conductance does not drop to $0$ due to the spin-splitting 
of the resonance condition, which can be recognized from the plot of conductance spin polarization 
$\Pol$ in \fig{Fano_vs_U1}(b). The latter varies continuously between $\Pol=-1$ 
(for $\delta_2$ corresponding  to the antiresonance in the majority spin channel) and $\Pol=1$ 
(for antiresonance in the minority channel). 
Qualitatively, this situation is hardly changed by finite Coulomb interactions 
in QD1 $U_1$ or the inter-dot capacitative coupling $U'$. 
It can be seen that $U_1$ slightly changes the position of the antiresonance and affects its width and depth.
On the other hand, $U'$ only shifts the minima, not affecting their depth or spin-splitting significantly, 
as can be seen from comparison with the $U'=0$ case,
which is plotted in \fig{Fano_vs_U1} with bright lines.

Basing on these observations, one could naively think that
a weak coupling of SC lead to QD1, effectively resulting in a reduction of $U_1$
to $\tilde{U}_1 = \sqrt{U_1^2 - 4\GS{1}^2}$,
should only quantitatively influence the Fano effect and its spin dependence.
As shall be shown in \sec{Fano_vs_SC1}, this conjecture is not true.

Summing up this section,
we have found that the presence of capacitive correlations between the two quantum dots
does not change qualitative features of the presented results.
However, the quantitative changes can be relatively strong,
due to exponential dependence of $T^*$ on $U'$.
Therefore, to make the analysis more realistic,
we assume $U' = U/10$ in our further analysis,
which is a reasonable value for typical experimental setups \cite{kellerNP14},
and discuss its influence on the results whenever important.

\section{Effect of pairing induced in the first quantum dot}
\label{sec:QD1-SC}

In this section we describe the properties of T-shaped DQD,
in which the first quantum dot is proximized by the superconductor, see \fig{system}(a).
In Sec.~\ref{sec:2Kondo_vs_SC1} we analyze how
the superconductor proximity affects the two-stage
Kondo effect in the considered system.
Then, in Sec.~\ref{sec:QPT_vs_t}, we examine the influence of the inter-dot hopping
on the phase transition in QD1 \cite{DomanskiIW}.
The interplay between the spin-dependent Fano interference
and the pairing induced by the SC lead is discussed in Sec.~\ref{sec:Fano_vs_SC1}.

\subsection{Influence of pairing correlations on the two-stage Kondo effect}
\label{sec:2Kondo_vs_SC1}

\begin{figure}
\includegraphics[width=0.9\linewidth]{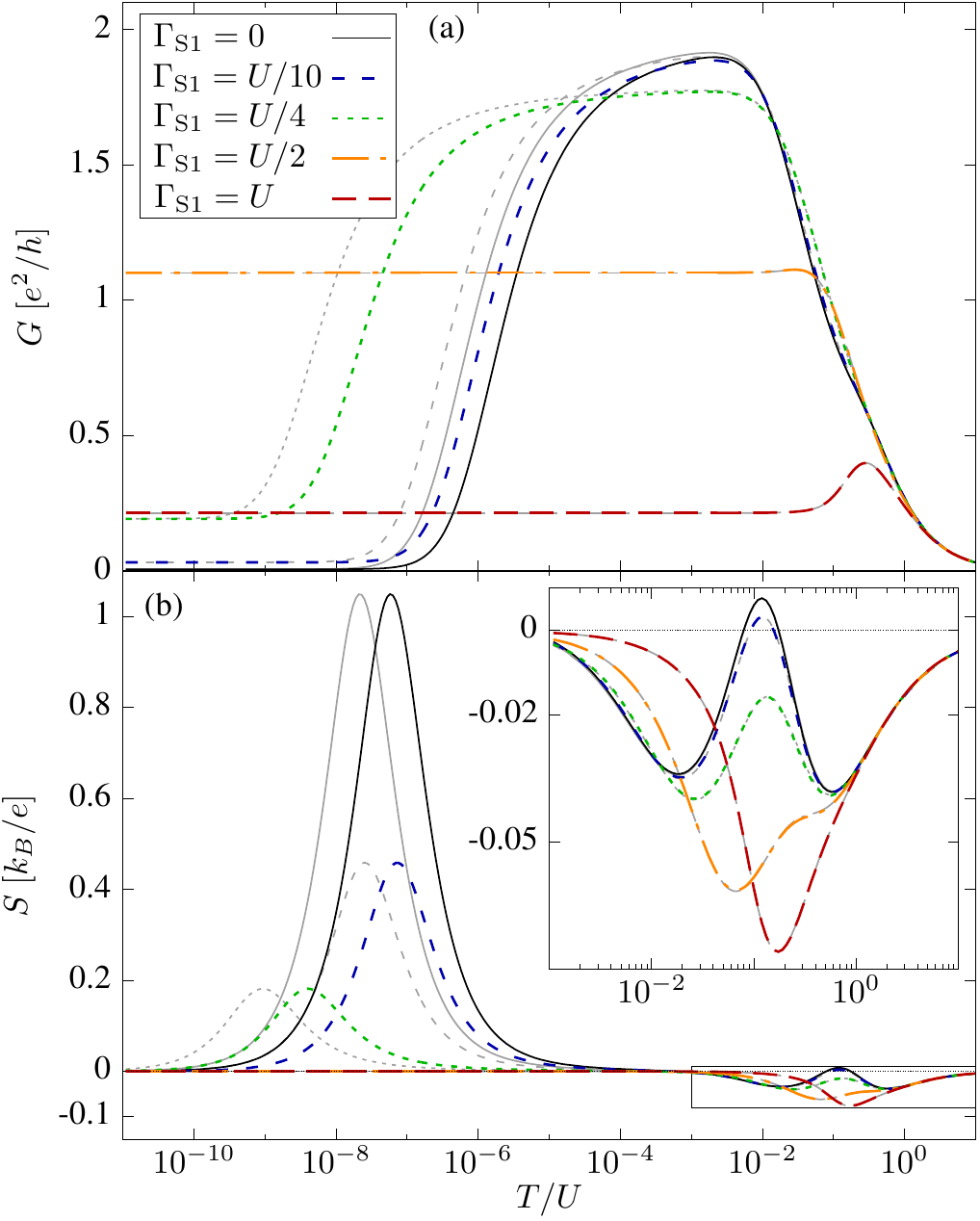}
\caption{(a) The conductance $G$ and (b) thermopower $S$ as a function of temperature $T$ 
		calculated for $U_i=U=D/10$, $U'=U/10$, and different values of coupling
		to superconducting lead $\Gamma_{{\rm S}1}$ at $\delta_1=0.05U$ and $p=0$.
		The case of $U'=0$ is shown with bright lines for comparison.
		The inset shows a close-up on the region of high $T$ in (b),
		marked by the rectangle in the main figure.
		}
\label{fig:Kondo_vs_SC1}
\end{figure}

\begin{figure*}
\includegraphics[width=0.85\linewidth]{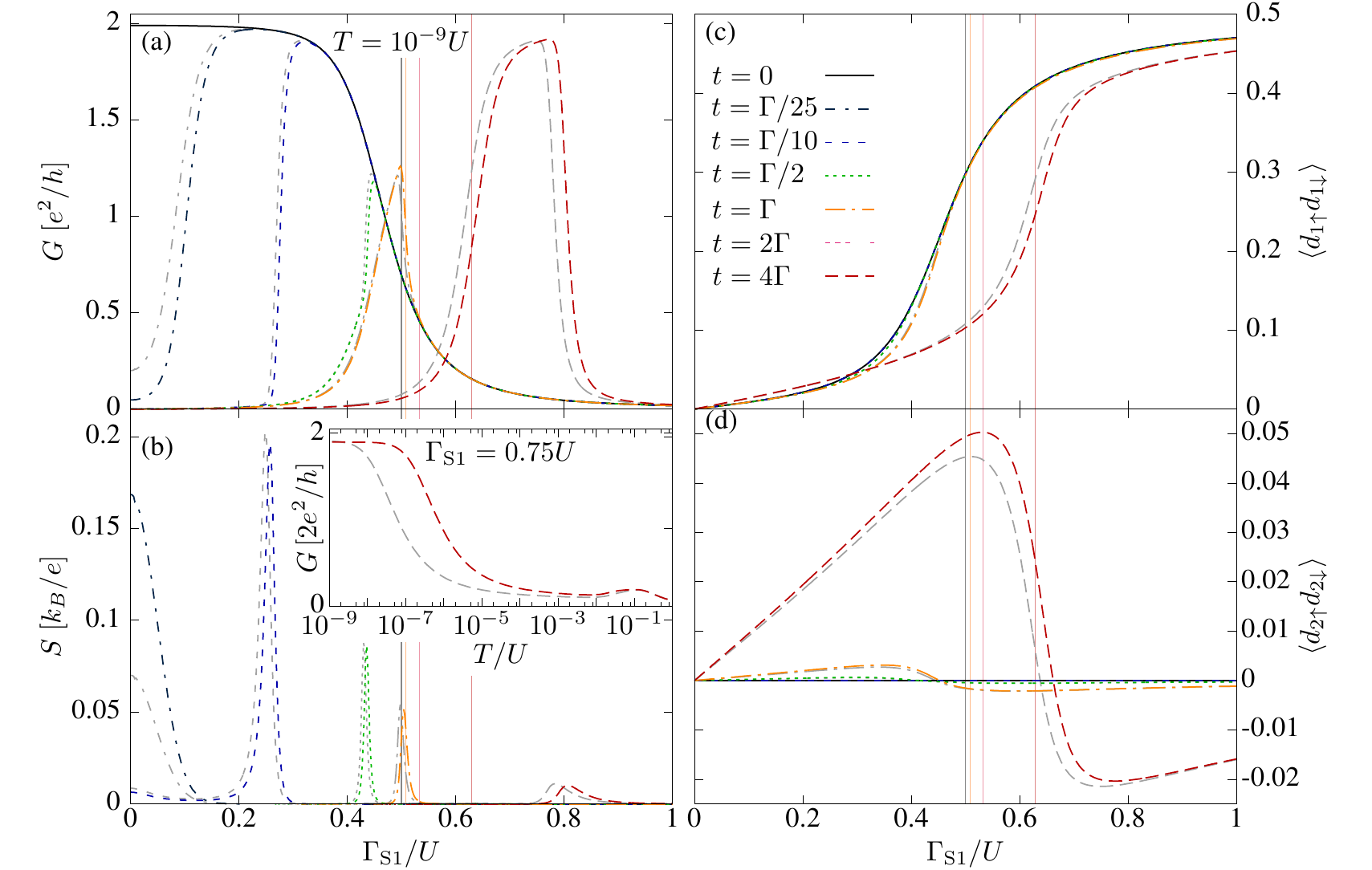}
\caption{(a) The conductance $G$, (b) the Seebeck coefficient $S$, and the expectation values
		(c) $\langle d_{1\up}d_{1\down}\rangle$ and (d) $\langle d_{2\up}d_{2\down}\rangle$
		 as functions of the coupling strength of the first quantum dot to the SC lead 
		 $\GS{1}$ for a few values of inter-dot hopping $t$, as indicated.
		 The other parameters are $\delta_1=0.05U$, for $U_1=U_2=U=D/10$, 
		 $T=10^{-9}U$, $\Gamma=U/20$, $p=0$ and $U'=U/10$.
		 The case of $U'=0$ is shown for comparison with bright lines.
		 The inset in (b) presents the temperature dependence of the conductance for 
		 $\GS{1}=0.75U$ and $t=4\Gamma$. Vertical lines correspond to the singlet-doublet transition
		 point for device decoupled from normal leads, corresponding to $t=0$ and $t\geq \Gamma$.
		}
\label{fig:QPT_vs_SC1}
\end{figure*}

The influence of the superconductor proximity on the two-stage Kondo effect
can be understood by resorting to the 
single-quantum-dot case, for which it was shown
that finite $\GS{1}$ ($\GS{2}=0$) 
results in an enhancement of the Kondo temperature \cite{DomanskiIW,KWIW}.
One can thus expect,
through exponential dependence of $T^*$ on $T_K$, 
cf. \eq{Tstar}, that even a small increase in $T_K$
should give rise to much larger changes in $T^*$.
This can be clearly seen in \fig{Kondo_vs_SC1}(a),
which presents the conductance plotted against $T$ for a few representative values of $\GS{1}$.
Indeed, while increasing the strength of coupling to the superconductor
results in a slight enhancement of $T_K$,
the second-stage Kondo temperature exhibits
a strong suppression with raising $\GS{1}$.
Additionally, for $\GS{1}<U/4$, one finds
$G(T=0) \approx \alpha \GS{1}^2/U^2$, with $\alpha \approx 3$.
Moreover, the local maximum in $G(T)$ is slightly lowered
as $\GS{1}$ increases. This can be understood
by referring to the case of a proximized quantum dot, 
where the low-temperature value of the conductance was found 
to be suppressed due to the coupling to superconductor \cite{DomanskiIW,KWIW}.
We also note that both the low-temperature conductance
as well as the local maximum in $G(T)$ 
are rather independent of $U'$,
although for $U'=0$ the minimum is achieved at slightly lower $T$,
due to smaller $T^*$, see \fig{Kondo_vs_SC1}(a).

Figure \ref{fig:Kondo_vs_SC1}(b) presents
how finite value of coupling $\GS{1}$ affects the thermopower of the system.
The most visible feature is that, unlike the conductance,
the Seebeck coefficient is very sensitive to the presence of SC correlations.
Already as small pairing potential as the one induced by $\GS{1}=U/10$ leads to
the reduction of maximal value of $S$ to less than a half of the value for $\GS{1}=0$.
One could claim that at low temperatures the thermopower
is proportional to $T$ and this reduction can be understood as a consequence
of decrease of $T^*$. However, usually the lower $T^*$ corresponds to the sharper dip in the 
spectral density, which compensates for the decrease of $T^*$.
In fact, when reducing the second-stage Kondo temperature $T^*$
by decreasing the hopping between the dots $t$,
the maximum in $S$ remains almost constant for $t<\Gamma/2$ \cite{KWIW-termo}.
Moreover, according to \fig{Kondo_vs_SC1}, 
the decrease caused by neglecting $U'$ also does not lead to the suppression of $S$, 
despite the fact that the corresponding decrease of $T^*$ is practically identical to 
the one caused by $\GS{1}=U/10$, cf.~\fig{Kondo_vs_SC1}(a).
One can conclude that the suppression of the thermopower by SC proximity effect cannot be explained
by the effective reduction of the Coulomb interactions and can be seen as a manifestation of the
sensitivity of caloric properties against the pairing correlations.

The values of thermopower at higher temperatures
are much smaller than at $T\sim T^*$, as already explained in \sec{Kondo_U}.
However, the zoom of $S$ in this regime (see the inset in \fig{Kondo_vs_SC1}) unveils
further interesting properties. First of all, as can be intuitively understood through
the effective reduction of $U_1$, the positive peak of $S(T)$ corresponding to the Coulomb 
blockade regime, is quickly suppressed with increasing $\GS{1}$.
Furthermore, the negative peak related to the Kondo regime
is enhanced and for strong $\GS{1}$ ultimately merges with the negative peak corresponding to the
thermal accessibility of the Hubbard peaks, see the curve for $\GS{1}=U$.
This behavior, clearly different compared to that for the second stage of screening,
shows that the competition between the SC correlations
and good thermoelectric properties is not a general rule.

\subsection{Influence of inter-dot hopping on the phase transition}
\label{sec:QPT_vs_t}

For negligible inter-dot hopping $t=0$, the system considered here is reduced to the case of 
a single quantum dot proximized by the SC lead, which has been studied, e.g. in Ref.~\cite{DomanskiIW},
in the context of the phase transition between the Kondo singlet and the singlet being 
a superposition of empty and doubly occupied states of the dot, where the expectation 
value $\langle d_{1\up}d_{1\down}\rangle$ becomes nonzero. 
This transition is a sharp quantum phase transition in the limit $\Gamma\to 0$ only,
while in the presence of normal leads it becomes a smooth crossover of the width $\sim \Gamma$.
In the following section we analyze the effect of finite hopping $t$ between the two dots on this 
crossover. To achieve this, we analyze the dependence of conductance, Seebeck coefficient and the
{\it order parameters} $\langle d_{1\up}d_{1\down}\rangle$ and $\langle d_{2\up}d_{2\down}\rangle$
as functions of the coupling to the SC lead, which are shown in \fig{QPT_vs_SC1}.
The coupling of QD1 to the normal leads $\Gamma$ was reduced in comparison to \fig{Kondo_vs_SC1}, 
to prevent the crossover from becoming very wide. In this way we can make reference to the physics 
of quantum phase transition, which only gets smeared due to finite $\Gamma$.
We also use cryogenic yet finite temperature, $T=10^{-9}U$, instead of $T=0$,
because for small values of $t$ the second-stage Kondo temperature
$T^*$ can be even smaller, which is experimentally completely irrelevant.

For $t=0$, the conductance smoothly changes from
almost $G=2e^2/h$ at $\GS{1}=0$ due to the conventional Kondo effect
(the value is slightly lower due to small detuning from PHS) to $G\approx0$
for strong $\GS{1}$, where the Kondo resonance at the
Fermi energy is destroyed by the pairing correlations.
As far as the thermopower is concerned, one could expect
a negative peak at $T\sim T_K$. However, at low temperatures $S\sim T$ \cite{CostiZlatic},
as follows from the Sommerfeld expansion, and for the considered very low temperature 
one gets $S \approx 0$.
The crossover of the order parameter at QD1 in the case of $t=0$ from $\langle d_{1\up}d_{1\down}\rangle = 0$,
in the absence of SC lead, to the universal limit $\langle d_{1\up}d_{1\down}\rangle = 1/2$, for
$\GS{1} \to \infty$, can be seen in \fig{QPT_vs_SC1}(c).
Obviously at the decoupled QD2, $\langle d_{2\up}d_{2\down}\rangle = 0$.
We note that the above discussed results are also valid
for finite $t$, as long as the hopping is such small that $T^* \ll T$.
Otherwise, the landscape changes significantly. 

For $\Gamma=U/20$, as assumed in \fig{QPT_vs_SC1},
finite value of hopping of the order of $t=\Gamma/25=U/500$ is already large enough to 
result in almost full development of the
second-stage of screening for $\GS{1}=0$ at the considered temperature.
However, finite $\GS{1}$ increases $T_K$ and consequently decreases $T^*$
[compare \eq{Tstar} and \fig{Kondo_vs_SC1}],
leading to the restoration of the conventional Kondo effect
(suppression of its second stage of screening) for some critical $\GS{1}$,
see the curves for $t=\Gamma/25$ and $t=\Gamma/10$ in \fig{QPT_vs_SC1}(a). 
This critical value of $\GS{1}^*$ corresponds to $T^*(\GS{1}=\GS{1}^*)=T$.
As explained in \sec{2Kondo_vs_SC1}, for $T\approx T^*$,
one can expect a large, positive peak in $S(T)$.
This condition is fulfilled around $\GS{1}=\GS{1}^*$
and, therefore, the corresponding peak of $S(\GS{1})$ can be
observed in~\fig{QPT_vs_SC1}(b). Again, for $t\leq U/40$ the couplings $\GS{1} \gtrsim U/2$ lead 
to the crossover to the Shiba state and the suppression of the Kondo effect,
with almost unaffected $\langle d_{1\up}d_{1\down}\rangle(\GS{1})$ dependence
and very small values of $\langle d_{2\up}d_{2\down} \rangle$.
In this sense, the crossover is qualitatively unaffected
by the presence of QD2, provided $t\ll \Gamma$.

Finally, let us analyze what happens for stronger values of hopping
$t\gtrsim \Gamma$. Then, for $\GS{1}=0$, the local singlet inside
the DQD is formed and the Kondo effect is completely suppressed \cite{Cornaglia}.
The transport is governed by the spectrum of $H_{\rm SDQD}$ and the matrix elements of $d_{1\s}$ between its 
eigenstates. When $\GS{1}$ is increased, at the critical value of $\GS{1}$, the ground state 
of $H_{\rm SDQD}$ becomes a spin doublet. In the limit of small $t$ this doublet corresponds
to a single electron in QD2 and QD1 in the superconducting singlet state. Therefore, the doublet 
is practically decoupled from the leads and the Kondo effect is suppressed.
However, inter-dot hybridization restores the matrix element of $d_{1\s}$ between the aforementioned doublet
and the excited states. Then, the Kondo effect is always present, although the corresponding Kondo
temperature $\tilde{T}_K$ vary strongly with $\GS{1}$. In particular, when the singlet-doublet 
splitting becomes very large, the relevant Kondo scale is strongly suppressed. This is visible 
in \fig{QPT_vs_SC1}(a) for $t=U/5$. On the other hand,
for $\GS{1} \sim 0.75U$, the Kondo effect is restored, as seen 
also in the inset, where the temperature dependence of conductance
for such a case is plotted. Higher values of 
$\GS{1}$ correspond to larger singlet-doublet splitting, hence the drop of $T_K$ below the 
temperature assumed for calculations in the figure.
We note that a similar suppression of the Kondo effect
due to singlet-doublet splitting was also reported
in the case of DQDs in a Cooper pair splitting geometry \cite{KacperIW}.

It seems worth emphasizing that the restoration of the Kondo effect for large $t$ does not have
the nature of suppressing the second stage of the Kondo effect. On the contrary, it happens rather
at QD2, while QD1 only mediates the coupling to the leads. This resembles the situation, when
QD1 is very far from particle-hole symmetry, described in Ref.~\cite{TanakaPRB2012}.
Interestingly, despite that the positive peak of $S(\GS{1})$ is only diminished, 
but not completely suppressed in this regime, although it does no longer coincide with maximum 
of $G(\GS{1})$ slope.
Moreover, for strong $t$, the order parameter at QD2 becomes nonzero; see \fig{QPT_vs_SC1}(d).
As long as the ground state of $H_{\rm SDQD}$ is a spin singlet, $\langle d_{2\up}d_{2\down}\rangle > 0$,
{\it i.e.} the order parameter in the second dot
has the same sign as $\langle d_{1\up}d_{1\down}\rangle$.
However, $\langle d_{2\up}d_{2\down}\rangle(\GS{1})$ changes sign at critical $\GS{1}$,
corresponding approximately to the singlet-doublet transition
in a DQD isolated from the normal leads. The critical values for the transition are indicated 
in \fig{QPT_vs_SC1} by vertical lines.
The sign change of the pairing expectation value
may be understood by recalling the fact that this is in fact
expected beyond the $\Delta\to \infty$ approximation,
i.e. when quasiparticle states in SC are also available \cite{OguriHewson}.
Since QD2 is proximized by the continuum of states formed
by QD1 and the leads, exhibiting also pairing correlations, the sign change
of its order parameter at the singlet-doublet transition is visible.
The difference between the zero of $\langle d_{2\up}d_{2\down}\rangle(\GS{1})$
and the value of $\GS{1}$ corresponding to the singlet-doublet
transition is a consequence of renormalization of DQD levels due to finite
coupling to normal leads $\Gamma$.

\subsection{Influence of pairing correlations on the spin-dependent Fano effect}
\label{sec:Fano_vs_SC1}

\begin{figure}
\includegraphics[width=0.9\linewidth]{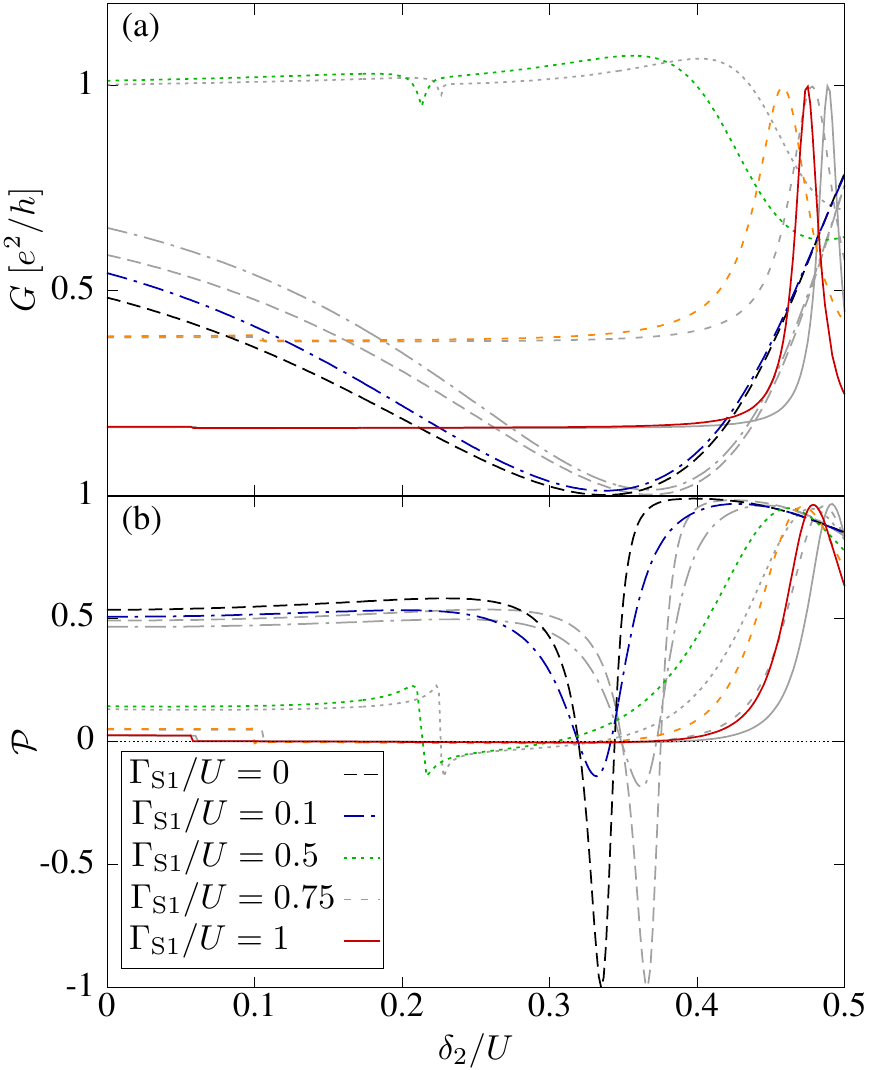}
\caption{(a) The low-temperature conductance $G$ and (b)                  
         its spin polarization $\Pol$ as a function 
		 of QD2 detuning $\delta_2$ for different couplings $\GS{1}$
		 between QD1 and SC lead, and $U_1=U_2=U=D/10$, $\Gamma=U/5$, $t=\delta_1 = U'=U/10$, $p=0.5$. 
		 Bright lines indicate the results in the case of $U'=0$, for comparison.
		 }
\label{fig:Fano_vs_SC1}
\end{figure}

From the discussion in previous sections, one can see that in many cases the 
main effect of the presence of a weakly coupled superconducting lead 
is an effective decrease of the relevant Coulomb interaction. However, this is not 
always the case, as argued in this section.
As shown in Sec.~\ref{sec:Fano_U}, 
in the case of ferromagnetic leads and $U_2 \neq 0$, the spin-dependent Fano effect 
is present irrespective of the Coulomb interaction strength in the first quantum dot, $U_1$.
Nevertheless, even relatively small values of $\GS{1}$ result in a practically complete 
suppression of the spin splitting of the minimum in conductance. 
This is visible in \fig{Fano_vs_SC1}, presenting the conductance and its spin polarization 
as functions of $\delta_2$ for $U_1=U_2=U$ and for a few representative values of $\GS{1}$. 
Although relatively low values of coupling $\GS{1}$ do not suppress the minimum 
in $G(\delta_2)$, see curve for $\GS{1}=0.1U$ in \fig{Fano_vs_SC1}(a), the spin filtering effect 
is completely suppressed, as presented in \fig{Fano_vs_SC1}(b). 
Note that such a suppression effect was not obtained by altering only $U_1$ 
in Sec.~\ref{sec:Fano_U}. Moreover, this effect does not depend on $U'$ either,
as can be seen by comparison to the case of $U'=0$ shown with bright lines in \fig{Fano_vs_SC1}.
The fragility of the spin-dependence of the Fano interference to the superconducting proximity effect 
is, therefore, a consequence of a nontrivial interplay between the pairing and the spin correlations.

In the case of stronger coupling $\GS{1}$, even more dramatic changes can be expected. 
Indeed, the Fano anti-resonance is completely removed
for $\GS{1} \geq 0.5 U$; see \fig{Fano_vs_SC1}(a).
Moreover, the transition between the singlet and 
doublet ground states of $H_{\rm SDQD}$
can give rise to the change of sign of the spin polarization, 
as observed in \fig{Fano_vs_SC1}(b);
see for example curve for $\GS{1}=0.5 U$ at $\delta_2 \approx 0.22U$.
Nevertheless, the suppression of conductance is not complete in any of spin channels and the
absolute value $|\Pol|$ does not exceed $25$\% in this regime.
One can thus conclude that superconducting pairing correlations have a clearly
detrimental effect on the spin filtering properties of the considered device.

\section{Effect of pairing induced in the second quantum dot}
\label{sec:QD2-SC}

In the preceding section the focal point of the discussion
was the phase transition in QD1 and its influence on the 
Kondo physics of the system.
Now, in turn, we move to the analysis of transport
properties of a different setup, which is shown in \fig{system}(b).
Even though the physics for small pairing correlations is in such a case quite similar
to the case of system presented in \fig{system}(a),
there appear significant differences which are discussed in the following. 

In the present section, the analysis of the Kondo effect is
continued for the case of small particle-hole asymmetry,
allowing for non-zero Seebeck coefficient to occur. The normal leads 
are assumed to be nonmagnetic. The Fano-like interference effects
occur to be very similar as in the case of pairing present in
QD1 and are not discussed in detail.
In particular, small values of $\GS{2}$ lead
to the Fano anti-resonance with suppressed spin-filtering effect,
while strong pairing correlations induced in the second quantum dot 
destroy the Fano effect completely. 

\subsection{Influence of pairing correlations on the two-stage Kondo effect}

\begin{figure}
\includegraphics[width=0.9\linewidth]{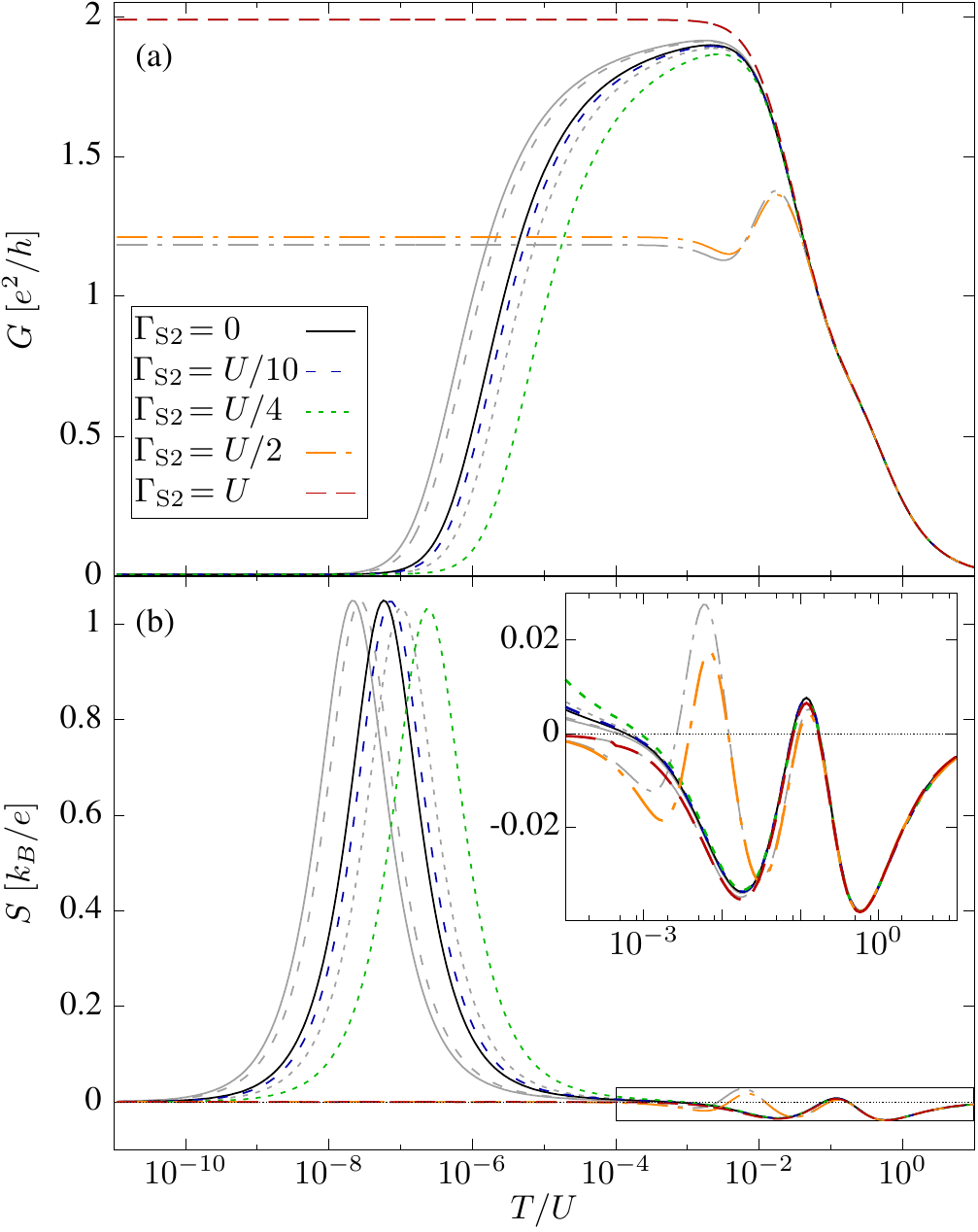}
\caption{
		(a) The conductance $G$ and (b) the Seebeck coefficient $S$ as functions of
		 the temperature $T$ for different couplings $\GS{2}$ between the second dot
		 and the SC lead calculated for $\delta_1=0.05 U$.
		 The other parameters are $U_1=U_2=U=D/10$, $\Gamma=U/5$, $t=\Gamma/4$, $p=0$, and $U'=U/10$. 
		 The case of $U'=0$ is shown with bright lines for comparison.
		 The inset shows the zoom into the large temperature region,
		 marked by rectangle in the main plot.
		}
\label{fig:Kondo_vs_SC2}
\end{figure}

For weak coupling between the second quantum dot and the SC lead,
$\GS{2} \ll U$, the qualitative understanding of the proximity effect
can be founded on the idea of effective reduction of $U_2$.
Therefore, the Kondo temperature for the first stage of screening
the spin in the first quantum dot, $T_K$, hardly depends on $\GS{2}$.
Furthermore, from \eq{Tstar} one immediately recognizes
that $T^*$ depends on $U_2$ through $J$, and grows with decreasing $U_2$.
Thus, for the device shown in \fig{system}(b), $T^*$ increases with $\GS{2}$ 
in a way similar to $T_K$ increasing with $\GS{1}$ for the one presented in \fig{system}(a).
Note that this is opposite to what happens to $T^*$ then.
This is illustrated in \fig{Kondo_vs_SC2}(a) for a few representative values of $\GS{2}$.
The corresponding change in the Seebeck coefficient peak
position can be observed in \fig{Kondo_vs_SC2}(b).

The physics changes, in comparison to pairing induced at QD1, for stronger interdot hopping $t$.
Here, the change of $H_{\rm SDQD}$ ground state corresponds
to the formation of a singlet in QD2, which suppresses the second 
stage of the Kondo effect for $\GS{2}$ above the critical value $\GS{2}^* \approx U/2$. This is
reflected in the perfect conductance and lack of the thermopower peak at low temperatures 
for $\GS{2} > \GS{2}^*$ (for $t>0$ and $\Gamma>0$ the transition is in fact a quite sharp crossover, 
as explained in the following subsection).
Interestingly, an additional sign change of $S(T)$ occurs at $T\sim T_K$ for $\GS{2}$
close to this critical value, as illustrated in the inset of \fig{Kondo_vs_SC2}(b).
This may be accounted for by the splitting of the Kondo peak by
a residual dip corresponding to the second stage of screening.
In fact, $T^*$ increases with $\GS{2}$ quite strongly and becomes only slightly smaller than $T_K$ 
for $\GS{2} \approx 0.4U$. Then, the slope of the QD1 spectral function at $\w=0$ changes
and implies the sign change of $S$.
Nevertheless, for even stronger $\GS{2}$, the second stage
of the Kondo effect becomes finally suppressed. 
Interestingly, the width of the dip in QD1 spectral density corresponding to the second stage 
of screening (which can be taken as a measure of $T^*$) is in fact still finite and even growing 
further, and only its depth vanishes, and so does the related positive peak of $S(T)$ together with 
the two corresponding sign changes.

\subsection{Phase transition in the second quantum dot}

\begin{figure}
\includegraphics[width=0.9\linewidth]{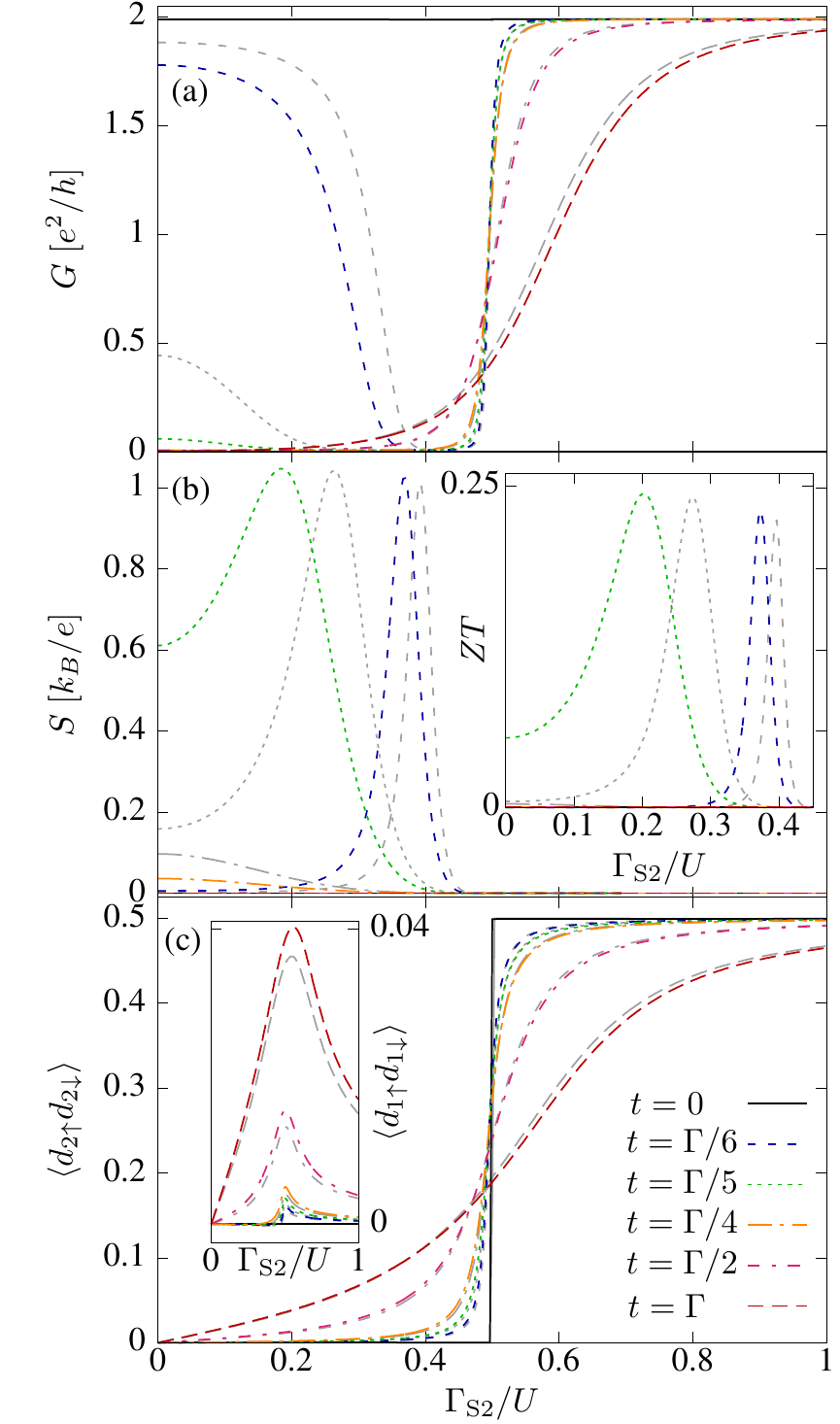}
\caption{
		 (a) The conductance $G$, (b) Seebeck coefficient $S$ and the order parameter in (c) QD2
		 $\langle d_{2\up} d_{2\down} \rangle$ as function of coupling between
		 the second quantum dot and the SC lead $\GS{2}$ for different inter-dot hoppings $t$.
		 The other parameters are $\delta_1 = 0.05U_1$,
		 $U_1=U_2=U=D/10$, $\Gamma = U/5$, $p=0$, $T=10^{-9}U$, and $U'=0.1U$.
		 The $U'=0$ case is shown with bright lines for comparison.
		 The inset in (b) shows the thermoelectric
		 figure of merit $ZT$ as a function of $\GS{2}$.
		 The inset in (c) presents the order 
		 parameter in QD1 (note different scale).
		}
\label{fig:QPT2_vs_t}
\end{figure}

The largest difference between the phase transition at QD1 and the one at QD2
induced by pairing correlations is associated with the fact that while QD1 is 
directly coupled to the metallic leads, QD2 is coupled only through QD1.
Therefore the effective broadening of QD2 levels
is in the leading order proportional to $\Gamma_2 \equiv t^2/\Gamma$. 
To explore the Kondo correlations one needs to consider relatively strong coupling $\Gamma$, which
leads to smearing of the transition at QD1.
On the contrary, the transition at QD2 is even sharper for strong $\Gamma$.
The effect is even more pronounced due to the fact that the interdot hopping $t$
in experimental setups can be quite small. Therefore, the crossover is in fact 
quite sharp and the similarity to the quantum phase transition, which occurs at 
$t=0$ or $\Gamma=0$, is even more evident than in the case of QD1. However, 
low values of $t$ also imply indeed cryogenic Kondo temperatures
for screening the second quantum dot spin, $T^*$, as follows from \eq{Tstar}.
This makes the system vulnerable to perturbations \cite{KWIW-stermo} 
and sets the ground for an interesting interplay between the Kondo effect
and the superconducting pairing correlations in the vicinity of the crossover region.

The main results concerning the influence of the interdot coupling on the
phase transition at QD2 are summarized in \fig{QPT2_vs_t}.
Similarly to \fig{QPT_vs_SC1}, a finite yet very small $T=10^{-9}U$ was assumed in calculations.
For $t=0$, there is a strict phase transition, with discontinuous change of the order parameter 
$\langle d_{2\up} d_{2\down} \rangle$ at $\GS{2}=U/2$, as shown in \fig{QPT2_vs_t}(c). At the same
time, there are no consequences of this fact for transport properties between the normal leads, 
since QD2 remains completely decoupled from them. Therefore, the conventional, single-stage Kondo
effect takes place and the conductance $G=G_{\rm max}$ does not depend on $\GS{2}$ ($G_{\rm max} < 
2e^2/h$ due to particle-hole asymmetry); cf. \fig{QPT2_vs_t}(a).
Similarly, the Seebeck coefficient $S\sim T \approx 0$, as shown in \fig{QPT2_vs_t}(b). 

For finite hopping $t$, the second stage of the Kondo effect develops at
energy scales corresponding to $T^*$.
Nevertheless, at finite temperature only for sufficiently strong $t$ does $T^*$
exceeds the actual $T$ used in calculations.
This can be visible for $t=\Gamma/6$ in \fig{QPT2_vs_t}(a).
Moreover, due to the increase of $T^*$ with $\GS{2}$,
the relevant critical value of $t$, at which $T^*=T$, diminishes.
Consequently, the conductance is suppressed and a peak appears
in $S(\GS{2})$ dependence; see \fig{QPT2_vs_t}(b).
However, unlike in the case of pairing induced in QD1 discussed in previous sections,
the obtained values of $S$ are larger and the thermoelectric efficiency is enhanced.
This is illustrated by the thermoelectric figure of merit
reaching almost $ZT=0.25$, as presented in the inset 
to \fig{QPT2_vs_t}(b). This should be compared to $ZT\approx 0.01$ for parameters assumed in
\fig{QPT_vs_SC1} (result not shown in the figure).
Further increase of the coupling to SC lead induces a crossover
to the conventional Kondo regime. Its width is set up by the
effective coupling of QD2 to the normal leads,
$\Gamma_2$, as can be deduced from \fig{QPT2_vs_t}(c).
Therefore, for strong $\GS{2}$, the conductance is maximized
and the thermopower strongly suppressed. 

It is interesting to note that in the geometry considered in this section,
QD2 and the normal leads do not form
a common continuous medium exhibiting pairing correlations,
to which QD1 is coupled.
For this reason, the pairing amplitude induced in QD1
by the coupling to QD2 is always of the same sign and is 
simply caused by the hybridization of single-electron states,
cf. inset in \fig{QPT2_vs_t}(c). 
Nevertheless, the order parameter $\langle d_{1\up} d_{1\down} \rangle$
exhibits a peak at $\GS{2} = \GS{2}^*$. 

Finally, the large $t$ regime corresponds to the transport through molecular levels of DQD
in the proximity of SC lead. The location of the crossover is only slightly shifted due to
the renormalization of the energy levels, but its width is increased significantly due to large 
$\Gamma_2$. As can be seen in the inset in \fig{QPT2_vs_t}(c), $\langle d_{1\up} d_{1\down} \rangle$ 
remains positive, which is due to the reasons explained above.
Consequently, the strong $t$ case
does not differ quantitatively much from the case corresponding to weaker inter-dot hopping, 
unless caloric properties are concerned. Then, of course, smoothed crossover
leads to a small slope of the spectral function at $\w=0$,
and consequently reduced thermopower.

\section{Conclusions}
\label{sec:conclusions}

In the present paper we have analyzed the transport properties
of a T-shaped double quantum dot system proximized by the superconductor,
considering two distinct geometries.
In the first one, the quantum dot directly coupled to the
normal leads was connected to the superconductor, while in the second geometry,
the side coupled quantum dot was proximized.
We have thoroughly examined the sub-gap physics
of both devices and showed that, depending on the 
superconductor position, the second-stage Kondo temperature
$T^*$ may be either enhanced or decreased by a small coupling to the superconductor.
In both cases there appears a doublet-singlet crossover
around some critical value of the SC pairing potential
and the properties of the system change completely for strong pairing correlations.
Depending on the device's geometry,
the conventional Kondo effect may be strongly supported
or completely suppressed in this transport regime.
Moreover, the crossover becomes very sharp for superconductor attached to 
side-coupled quantum dot at the regime of strong coupling to normal leads.
We explain these effects as consequences of effective decrease
of the corresponding Coulomb interaction and basic properties of coupled Kondo impurities.
Moreover, we show that the spin-dependent Fano-Kondo 
interference, which develops in the considered systems,
turns out to be very vulnerable to the proximity effect.
The spin-filtering effects present in T-shaped DQDs with ferromagnetic contacts
can be suppressed by even small values of the coupling to the superconductor.

The presented results show that the superconductor proximity effect
provides additional means for the control of the two-stage Kondo physics
in T-shaped double quantum dots. It enables to either strongly favor 
or completely suppress each stage of the Kondo screening
and obtain interesting electric or thermoelectric properties.
Furthermore, the analysis of transport properties of hybrid T-shaped DQD systems
gives additional insight into the nature of the interplay between the
Kondo correlations and the superconductivity, which exhibits 
a surprising combination of increase of the Kondo temperature 
and suppression of the related spectral features.
We hope that our analysis will foster further endeavor in this direction.

\acknowledgments

This work was supported by the National Science Centre in Poland through project no.
2015/19/N/ST3/01030. Discussions with B. Bu\l{}ka are acknowledged.


\end{document}